\documentclass[reprint, a4paper,superscriptaddress,floatfix,amsmath,amssymb,amsfonts,noshowpacs,]{revtex4-1}
\pdfoutput=1 
\usepackage{newtxtext,newtxmath}
\usepackage[utf8]{inputenx} 
\usepackage[dvipsnames]{xcolor}
\usepackage{soul} 
\usepackage[textwidth=17.5cm,textheight=23.5cm,verbose,pdftex]{geometry}
\usepackage{floatrow}
\usepackage[pdftex]{hyperref}
\usepackage{enumerate}
\usepackage{amsmath,amsfonts,amssymb} 
\usepackage{booktabs} 
\usepackage{multirow} 

\hypersetup{pdfauthor={Thomas Auzelle}, bookmarksnumbered=true, pdftitle={Electronic properties of air-exposed GaN$(1\bar{1}00)$ and $(0001)$ surfaces after several device processing compatible cleaning steps}, colorlinks, citecolor=blue, linkcolor=blue, urlcolor=blue}

\usepackage{gensymb}
\usepackage{graphicx}

\newcommand{\etal}[1]{\textit{et al.} [\onlinecite{#1}]}
\newcommand{\ie}{\textit{i.e.}}
\newcommand{\eg}{\textit{e.g.}}
\newcommand{\degC}{\,$^\circ$C}

\newcommand{\insitu}{\textit{in~situ}}

\begin{document}
\title{Electronic properties of air-exposed GaN$(1\bar{1}00)$ and $(0001)$ surfaces after several device processing compatible cleaning steps}

\author{T. Auzelle} \email[Electronic mail: ]{auzelle@pdi-berlin.de}
\affiliation{Paul-Drude-Institut für Festkörperelektronik, Leibniz-Institut im Forschungsverbund Berlin e.\,V., Hausvogteiplatz 5--7, 10117 Berlin, Germany}

\author{F.~Ullrich}
\affiliation{InnovationLab, Speyerer Str. 4, 69115 Heidelberg, Germany}
\affiliation{Materials Science Department, Technische Universit\"at Darmstadt, Otto-Berndt-Strasse 3, 64287 Darmstadt, Germany}

\author{S.~Hietzschold}
\affiliation{InnovationLab, Speyerer Str. 4, 69115 Heidelberg, Germany}
\affiliation{Institute for High-Frequency Technology, Technische Universit\"at Braunschweig, Schleinitzstrasse 22, 38106 Braunschweig, Germany}

\author{S.~Brackmann}
\affiliation{InnovationLab, Speyerer Str. 4, 69115 Heidelberg, Germany}

\author{S.~Hillebrandt}
\affiliation{InnovationLab, Speyerer Str. 4, 69115 Heidelberg, Germany}
\affiliation{Kirchhoff Institute for Physics, Heidelberg University, Im Neuenheimer Feld 227, 69120 Heidelberg, Germany}


\author{W.~Kowalsky}
\affiliation{InnovationLab, Speyerer Str. 4, 69115 Heidelberg, Germany}
\affiliation{Institute for High-Frequency Technology, Technische Universit\"at Braunschweig, Schleinitzstrasse 22, 38106 Braunschweig, Germany}
\affiliation{Kirchhoff Institute for Physics, Heidelberg University, Im Neuenheimer Feld 227, 69120 Heidelberg, Germany}


\author{E.~Mankel}
\affiliation{InnovationLab, Speyerer Str. 4, 69115 Heidelberg, Germany}
\affiliation{Materials Science Department, Technische Universit\"at Darmstadt, Otto-Berndt-Strasse 3, 64287 Darmstadt, Germany}

\author{R.~Lovrincic}
\affiliation{InnovationLab, Speyerer Str. 4, 69115 Heidelberg, Germany}
\affiliation{Institute for High-Frequency Technology, Technische Universit\"at Braunschweig, Schleinitzstrasse 22, 38106 Braunschweig, Germany}

\author{S.~Fern\'andez-Garrido} \email[Electronic mail: ]{sergio.fernandezg@uam.es}
\affiliation{Paul-Drude-Institut für Festkörperelektronik, Leibniz-Institut im Forschungsverbund Berlin e.\,V., Hausvogteiplatz 5--7, 10117 Berlin, Germany}
\affiliation{Grupo de Electr\'onica y Semiconductores, Dpto. F\'isica Aplicada, Universidad Aut\'onoma de Madrid, C/ Francisco Tom\'as y Valiente 7, 28049 Madrid, Spain}

\begin{abstract}
We report on the electronic properties of GaN$(1\bar{1}00)$ and $(0001)$ surfaces after three different and subsequent device processing compatible cleaning steps: HCl etching, annealing at $400$\degC{} in N$_2$ atmosphere, and O$_2$ plasma exposure. The surface electronic properties are quantified, in the dark and under ultraviolet illumination, using X-ray photoelectron spectroscopy and a Kelvin probe. We find that the cleaning steps largely affect the work function and the band bending of both GaN orientations. These modifications are attributed to the presence of different surface states as well as to the formation of adsorbates building up distinct surface dipoles. 
Besides these results, we detect that under ultraviolet illumination the work function of the surfaces exposed to HCl decreases by at least $0.2$\,eV without screening of the band bending. We thus attribute the observed surface photovoltage to a photo-induced modification of the surface dipole. Overall, these results emphasize the strong dependence of the electronic properties of air-exposed GaN surfaces on adsorbates. As a result, we advocate the use of the common cleaning steps analyzed here to re-initialize at will GaN$(1\bar{1}00)$ and $(0001)$ surfaces into pre-defined states.
\end{abstract}

\keywords{
GaN surface, cleaning, surface photovoltage, nonpolar orientation
}

\maketitle

\textcopyright 2019. This manuscript version is made available under the \href{http://creativecommons.org/licenses/by-nc-nd/4.0/}{CC-BY-NC-ND 4.0 license}.

\section{Introduction}
For GaN surfaces exposed to air, ensuring reproducible electronic properties through time and experiments is a basic but non-trivial requirement. 
A clear example is the photoluminescence intensity of air-exposed GaN$(0001)$ films, which has been reported either to increase \cite{Martinez2000,Chen2007} or decrease \cite{Shalish2001,Tripathy2002} after an HCl treatment. This discrepancy evidences different initial states for air-exposed surfaces, owing to the substantial sensitivity of their electronic properties to adsorbates \cite{Foussekis2009,Pfuller2010,Gurwitz2011,Sanford2013,Posada2013,Kim2014,Janicki2016,Wang2017}, specific preparation methods \cite{Bartos2016}, ultraviolet (UV) light exposure \cite{Behm2000,Foussekis2009,Pfuller2010,Hetzl2018}, and air-oxidation conditions \cite{Dong2006a,Dong2006b}. To overcome these issues, surface cleaning can be used to re-initialize at will air-exposed surfaces into a predefined state. A wealth of cleaning methods has been developed, as reported in the extensive and detailed article of Bermudez \cite{Bermudez2017}. The author concludes by advocating the use of cleaning steps performed \insitu{} under high vacuum conditions -- annealing above $500$\,$^\circ$C under Ga or NH$_3$ fluxes and/or N ion sputtering -- to reproducibly obtain GaN surfaces free of adsorbates and with a well-defined atomic order. The feasibility of such approach, which requires handling the cleaned samples in vacuum, was confirmed by Barto\v{s} \etal{Bartos2016}. Alternatively, if limiting ourselves for practical reasons to surface treatments not done in high vacuum, Bermudez's article reviews a number of common cleaning steps (\eg{}, acid etching, UV/O$_3$ exposure) that systematically ends up in GaN surfaces with similar levels of adsorbates. Although these simple cleaning steps do not result in as well-defined surfaces as those obtained using high-vacuum cleaning methods, they remain better suited for the fabrication and analysis of devices, which are usually processed, characterized, and/or operated in air. Furthermore, these common cleaning steps are equally applicable for treating thin films and three-dimensional nanostructures, such as nanowires and fins. These alternative cleaning methods are thus of extreme practical interest but require a proper characterization of the electronic properties of the resulting \textit{not ideal} GaN surfaces.

In this letter, we report on the electronic properties of non-intentionally doped n-type GaN$(1\bar{1}00)$ and $(0001)$ surfaces after three common cleaning steps, namely, HCl etching, annealing in N$_2$ atmosphere, and dry oxidation by O$_2$ plasma exposure. Absolute values for both the band bending (BB) and the work function (WF) are measured in the dark and under UV-A light illumination using X-ray photoelectron spectroscopy (XPS) and a Kelvin probe. Our results reveal that the BB and the WF strongly depend on the previous cleaning steps due to the presence of different residual adsorbates, which modify the surface states and introduce distinct surface dipoles.

\section{Material and methods}
The investigated GaN$(1\bar{1}00)$ surfaces are obtained from $5\times10$\,mm$^2$ free-standing GaN$(1\bar{1}00)$ substrates with a $1^\circ$ off-cut toward the $[000\bar{1}]$ direction. These substrates, purchased from Suzhou Nanowin Science and Technology, exhibit a low density of threading dislocations ($< 5 \times 10^{6}$\,cm$^{-2}$) and a root mean square surface roughness below $0.2$\,nm. The surface of the as-received substrates, prepared by the vendor using chemical-mechanical polishing, does not exhibit atomic terraces. To improve the atomic order at the surface, we perform a $\approx1$\,$\mu$m GaN regrowth by plasma-assisted molecular beam epitaxy. %
To preserve the surface smoothness, the growth is carried out at $\approx700$\degC{} under intermediate Ga-rich conditions \cite{Shao2013,Lim2017}.
The analyzed GaN$(0001)$ surfaces are obtained from a commercial 2$^{\prime\prime}$ GaN template purchased from Kyma Technologies. Such GaN template consists of a $\approx5$\,$\mu$m thick GaN layer epitaxially grown on Al$_{2}$O$_{3}$(0001) by hydride vapor phase epitaxy. The nominal threading dislocation density of the GaN layer is $< 1 \times 10^9$\,cm$^{-2}$. The investigated GaN$(1\bar{1}00)$ and $(0001)$ layers are both non-intentionally doped but exhibit n-type conductivity with a concentration of donors minus acceptors in the upper $10^{16}$\,at/cm$^{3}$ range, as derived from capacitance voltage profiling using a two-point Hg-probe. 

The three consecutive cleaning steps subject of the present study are:
\begin{itemize}
	\item[1.] \textit{HCl etching}: the GaN sample is dipped for $10$\,min in a $32\%$ HCl ($<0.0012\%$ contaminants) aqueous solution and subsequently rinsed in a boiling $99\%$ pure ethylacetate solvent.
	\item[2.] \textit{Annealing}: the GaN sample is annealed at $400$--$500$\degC{} for a minimum of $20$\,min in a N$_2$ atmosphere.
	\item[3.] \textit{O$_2$ plasma}: the GaN sample is exposed for $5$\,min to an O$_2$ plasma (forward power of $900$\,W in a O$_2$ atmosphere of $0.3$\,mbar), which is conceptually similar to UV/O$_3$ exposure, although a larger kinetic energy is provided to O radicals. 
\end{itemize}
As largely reported for the GaN$(0001)$ surface, \textit{HCl etching} decreases to low levels the areal concentration of typical surface contaminants (C and O) [Ref. \cite{Bermudez2017}, section 3.1] and prevents reoxidation \cite{Uhlrich2008}. The subsequent \textit{annealing} step reduces the amount of residual Cl left after \textit{HCl etching} \cite{Rickert2002,Uhlrich2008}. At last, the \textit{O$_2$ plasma} step efficiently removes carbonaceous compounds [Ref. \cite{Bermudez2017}, section 3.1] and creates a capping GaO$_x$ layer with an estimated thickness $\lesssim1$\,nm \cite{Irokawa2017,Irokawa2018}. The formation of a much thicker GaO$_x$ layer is further excluded here, using reflection high-energy electron diffraction, by the observation of a $1\times1$ reconstructed GaN$(0001)$ surface. In the following, for the sake of simplicity, we treat this GaO$_x$ capping as one interface (GaN/ambient) and not as an heterostructure with two interfaces (GaN/GaO$_x$/ambient).

The electronic properties of the GaN surfaces are probed by XPS in a PHI Versa Probe II equipped with a monochromatized Al K$_{\alpha}$ anode as X-ray source ($1486.6$\,eV). The base pressure in the XPS chamber is in the $10^{-9}$\,mbar range. The kinetic energy of photo-emitted electrons is measured using a concentric hemispherical analyzer. The takeoff angle is $90^\circ$ and the angle between the analyzer and the X-ray source $54.7^\circ$. All spectra are referenced in binding energy with respect to the Fermi edge of Ag. The secondary electron edges are calibrated according to the WF of an \textit{in situ} sputter-cleaned Ag foil. The WF of the Ag foil is determined by ultraviolet photoelectron spectroscopy. GaN WFs are then assessed as $WF = h\nu_{K \alpha} - BE_{SEC}$, where $h\nu_{K \alpha}$ is the energy of the X-ray source and BE$_{SEC}$ the binding energy of the secondary electron cutoff.
XPS fits are performed using Voigt line profiles after subtraction of a Shirley background \cite{Shirley1972}. Measurements are performed in the dark or under the illumination of a focused high-brightness light-emitting diode (LED) with $\lambda=365$\,nm (above band gap excitation) resulting in an approximate excitation density of $10$--$100$\,mW/cm$^2$ at the sample surface. Importantly, for our XPS measurements, photoemission-induced screening of the GaN surface BB \cite{Long2002} has a negligible effect, as deduced from Fermi edge measurements of $3$--$4$\,nm thick Au layers evaporated on uncleaned and \textit{O$_2$ plasma} exposed GaN$(0001)$ surfaces. The position of the Fermi level is systematically found at $0.00 \pm 0.03$\,eV, indicating the absence of GaN BB screening. 

Complementary measurements of the surface WF are done by Kelvin probe force microscopy (KPFM) in a N$_2$ filled environmental chamber. A Bruker Multimode 8 setup is used with the amplitude-modulated two-pass lift-off configuration. It is chosen to favor the WF resolution over the spatial resolution. The first pass provides the topographic measurements and the second one the contact potential difference (CPD) between the oscillating cantilever and the surface. A lift-height of $\approx 500$\,nm is used during the CPD measurement to minimize the contribution from the apex of the probing tip \cite{Colchero2001}, which is prone to be contaminated or damaged during measurements. We use a platinum-iridium coated Si probe, whose WF is calibrated before and after the measurements against freshly cleaved highly oriented pyrolytic graphite (WF of $4.5 \pm 0.1$\,eV \cite{Hansen2001}). Measurements are performed in the dark or under the illumination of a LED emitting at $405$\,nm (sub-band gap excitation) with an excitation density in the order of $0.1$\,mW/cm$^2$ at the sample surface. We note that for these complementary measurements, a different set of GaN samples has been used as well as a different lab equipment for the three cleaning steps.

\section{Results and discussion}
\subsection{Surface morphology and adsorbates}
\begin{figure}
\includegraphics[width = \linewidth]{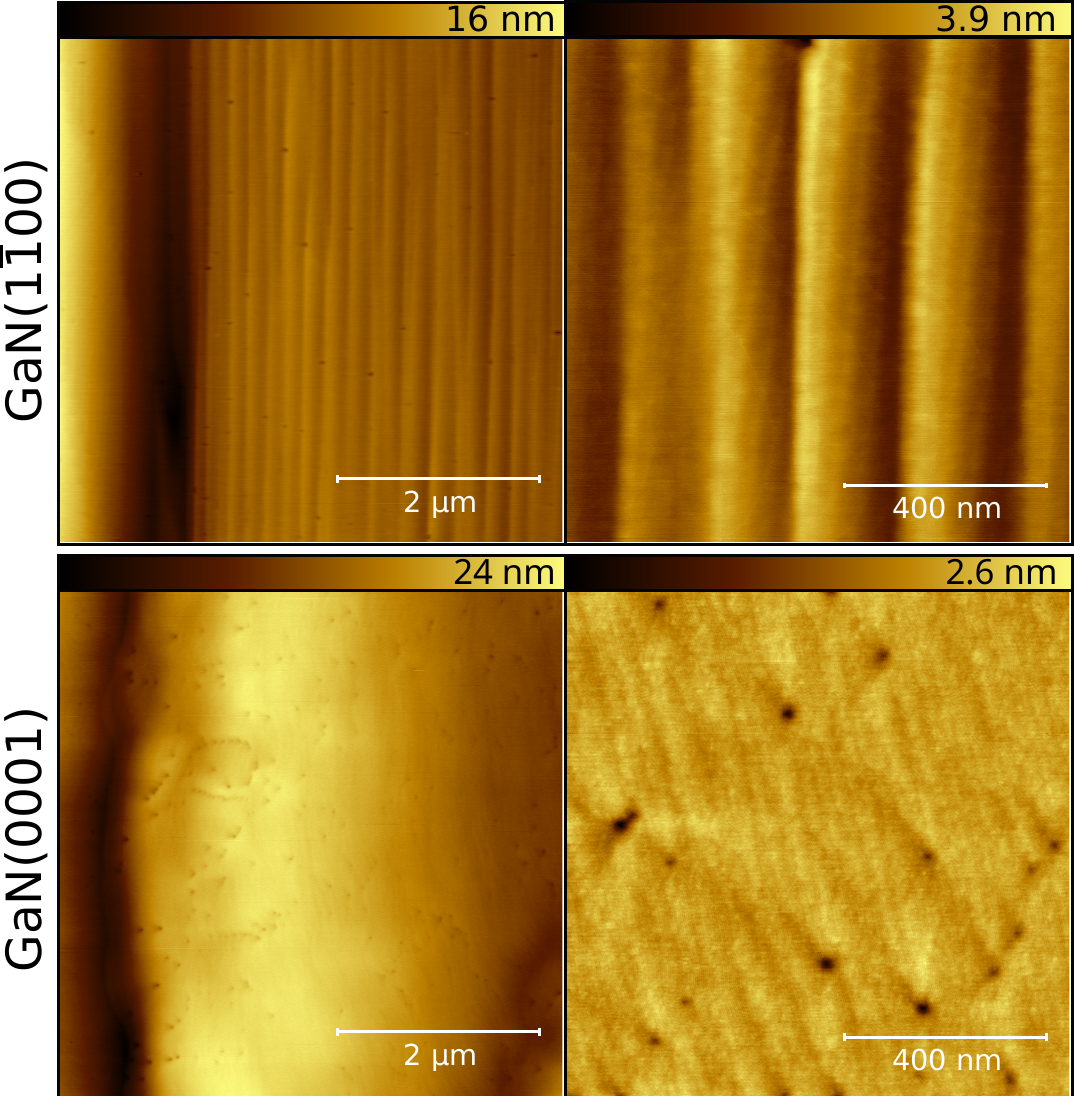}%
\caption{Representative atomic force micrographs with different magnifications of the GaN$(1\bar{1}00)$ and $(0001)$ surfaces under scrutiny.
\label{fig:AFM}}
\end{figure}

The surface morphology of the as-grown GaN$(1\bar{1}00)$ and as-received GaN$(0001)$ samples is characterized by atomic force microscopy (AFM). Representative micrographs of both samples are shown in Fig. \ref{fig:AFM}. The two different surfaces exhibit atomic terraces, which we use as a hallmark for a well-defined surface. Additional features such as dislocation related pits and step bunching are further neglected due to their small footprint on the surface. 

The amount of extrinsic atoms (C, O and Cl) remaining on the GaN surfaces after each of the three cleaning steps is evaluated by XPS. The integrated intensity of the O\,$1s$, C\,$1s$ and Cl\,$2p$ core-levels are gathered in Table\,\ref{tab:sumup} after correction with the relative sensitivity factors of Ref.\,\cite{Scofield1976}. We notice that all samples were exposed beforehand to an \textit{O$_2$ plasma} to initialize them. The cleaning steps are seen to act similarly on the GaN$(1\bar{1}00)$ and $(0001)$ surfaces. After \textit{HCl etching}, surface C, O and Cl atoms are observed, likely forming $\cdot$ClO$_n$H$_m$, $\cdot$CO$_n$H$_m$, $\cdot$OH$_n$ and $\cdot$H terminations. Subsequently to the \textit{annealing} in a N$_2$ atmosphere, the Cl atoms are gone and only C, O and H atoms remain chemically bound to the GaN surface. Finally, after the \textit{O$_2$ plasma}, only O atoms subsist on the surface, likely forming a thin GaO$_x$ layer with $\cdot$OH$_n$ terminations.

\begin{table*}
\caption{Relative adsorbates surface areal density, surface BB, WF and surface dipole ($\delta$) extracted from XPS and KPFM measurements. If not mentioned, the last significant digit indicates the measurement precision.
\label{tab:sumup}}
\begin{tabular}{cc|ccc|ccccc|cc}
	\toprule
	Surface											& Process 		& \multicolumn{3}{c}{Relative adsorbate}	& \multicolumn{5}{c}{XPS surface analysis}													& \multicolumn{2}{c}{KPFM analysis}	\\
		orientation								&							& \multicolumn{3}{c}{density (arb. units)}	& \multicolumn{5}{c}{(eV)}																										& \multicolumn{2}{c}{($\pm 0.05$\,eV)}	\\
	\midrule

    &							& O				& C 					& Cl 				& Ga $2p_{3/2}$				& BB 			& WF$_\text{dark}$ 			& $\delta$ & $\Delta \delta_\text{LED}^{365\,nm}$ 			& WF$_\text{dark}$	& $\Delta \delta_\text{LED}^{405\,nm}$\\
	&&&&&&&&&&&\\
\multirow{3}{*}{$(1\bar{1}00)$}&HCl					& 4		& 15				& 14			&	1118.99							& -0.2		& 3.9			&0.1				& -0.2										& 4.45				& -0.20			\\
															&N$_2$ annealing& 10		& 18				& <1			& 1118.76							& 0.1			& 4.0		&-0.1					& 0.0											& 4.30				& -0.05			\\
															&O$_2$ plasma	& 150		& <1				& <1			& 1118.30							& 0.5			& 5.3		&0.7					& 0.0										& 5.10				& -0.05			\\
	&&&&&&&&&&\\
\multirow{3}{*}{$(0001)$} 		&HCl					& 12		& 3					& 10			&	1118.83							& 0.0				& 4.8			& 0.8				& -0.2										& 4.55				& -0.35			\\
															&N$_2$ annealing& 21		& 36				& 1			&	1118.58							& 0.2			& 3.7	&-0.1						& 0.0											& 4.25				& -0.05			\\
															&O$_2$ plasma	& 63		& <1				& <1			&	1118.24							& 0.6			& 5.4		&0.8					& 0.0										& 4.90				& -0.10				\\
	\bottomrule
						
\end{tabular}%
\end{table*}
%
\begin{figure*}
\includegraphics[width = .8\linewidth]{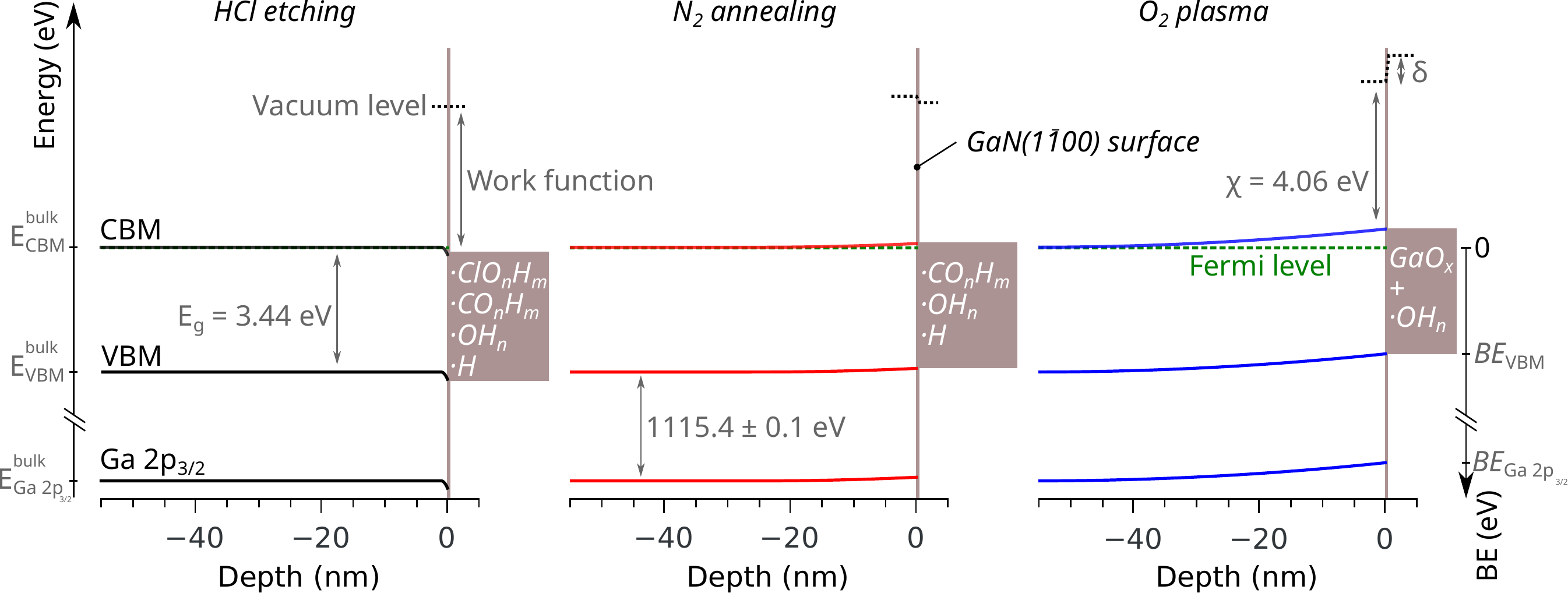}%
\caption{On scale GaN$(1\bar{1}00)$ surface band structure measured in vacuum and in the dark by XPS after the three consecutive cleaning steps. $\delta$ refers to the surface dipole and $\chi$ to the electron affinity for ``bare'' GaN, chosen equal to $4.06$\,eV\,\cite{Portz2018}. In the case of \textit{HCl etching}, we assume a constant electron density of $1 \times 10^{20}$\,cm$^{-3}$ in the CB, in analogy with the InN case in Ref.\,\cite{Mahboob2004}.
\label{fig:Diagram}}
\end{figure*}

\begin{figure*}
\includegraphics[width = .7\linewidth]{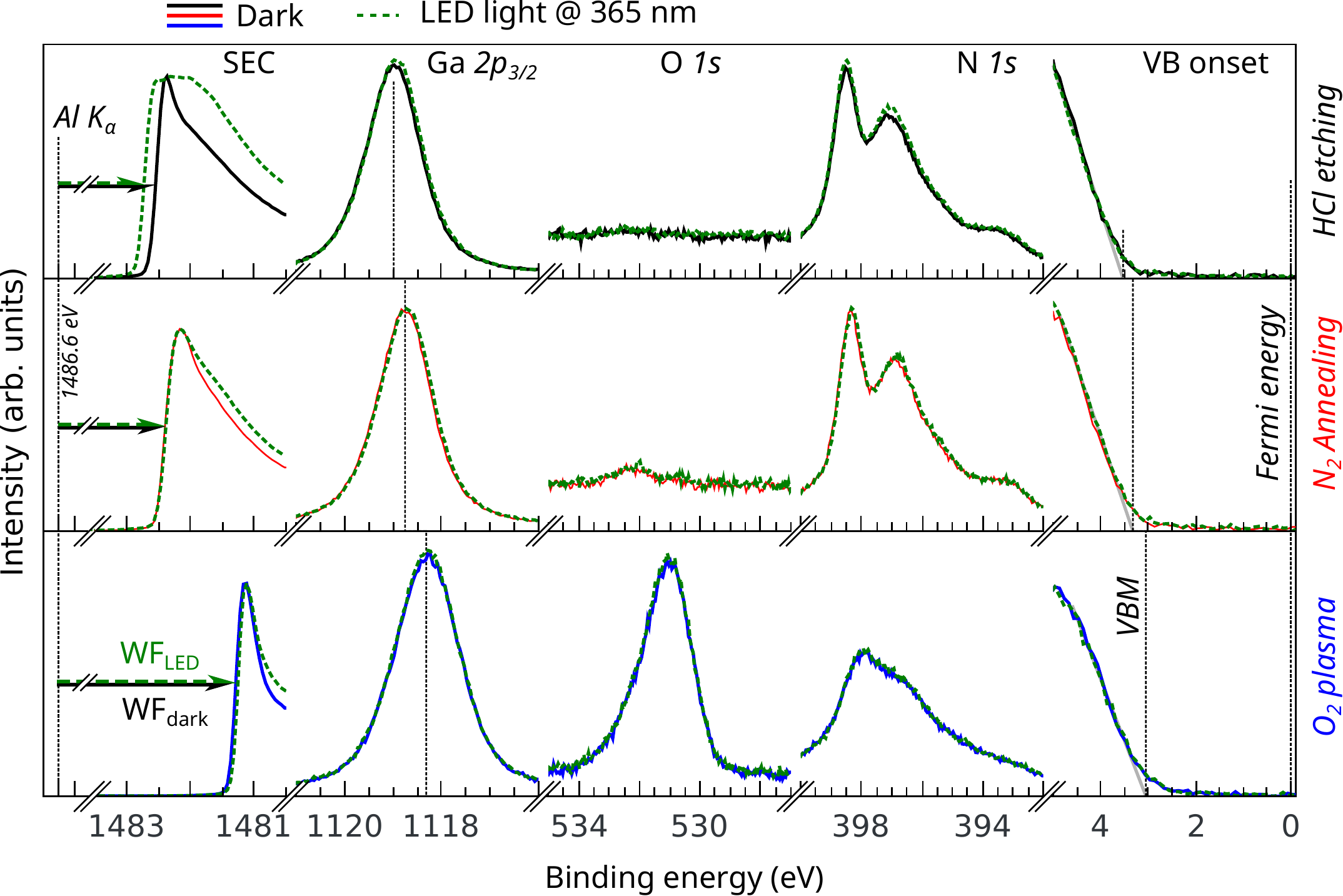}%
\caption{XPS detailed spectra acquired in the dark and under UV-A illumination of the valence band maximum (VBM), the N\,$1s$, the O\,$1s$, the Ga\,$2p_{3/2}$, and secondary electron cutoff (SEC) regions for the GaN$(1\bar{1}00)$ surface after \textit{HCl etching}. Vertical dotted lines indicate the energy of the VBM and Ga\,$2p_{3/2}$ as deduced by the line shape fitting.
\label{fig:ligthdependence}}
\end{figure*}
\subsection{Band bending}
XPS is further used to quantify the GaN surface BB in vacuum after each cleaning step. The surface BB can be directly extracted from the binding energy (BE) of the valence band maximum (VBM), although with a mediocre precision due to the difficulty to accurately fit the valence band onset \cite{Waldrop1996}. Instead, similarly to the method advocated by Bermudez \cite{Bermudez2017}, we measure the binding energy of the Ga $2p_{3/2}$ core level ($BE_{\text{Ga }2p3/2} $) and relate it to the surface BB through: 
\begin{equation}
\begin{aligned}
    BB =& [E_\text{CBM} - E_F]_{surface} - [E_\text{CBM} - E_F]_{bulk} \\
    BB =& [E_\text{VBM} - E_{\text{Ga }2p3/2}] + E_g - BE_{\text{Ga }2p3/2} \\
	&- [E_\text{CBM} - E_F]_{bulk} \\
\end{aligned}
\end{equation}
where $E_\text{CBM}$ is the conduction band minimum (CBM) energy, $E_\text{VBM}$ the VBM energy, $E_F$ the Fermi energy, and $E_g$ the electronic band gap. These quantities are indicated on the sketched GaN surface band structures depicted in Fig.~\ref{fig:Diagram}. The value $[E_\text{VBM} - E_{\text{Ga }2p3/2}]$ is a material constant and equals $[BE_{\text{Ga }2p3/2} - BE_\text{VBM}]$. By averaging the value extracted for $25$ XPS spectra of GaN samples with different orientations and surface adsorbates, we obtain $[E_\text{VBM} - E_{\text{Ga }2p3/2}] = 1115.4 \pm 0.1$\,eV. The error bar relates to the uncertainty in the determination of the VBM binding energy (a simple linear fit of the valence band leading edge). At room temperature, unstrained n-type GaN ($N_d \approx 10^{17}$\,at/cm$^{3}$) has $E_g = 3.44$\,eV \cite{Feneberg2014} and we assume the quantity $[E_\text{CBM} - E_F]_{bulk}$ nearly equal to the dopant ionization energy, which in our case amounts to $20\pm10$\,meV for Si and O dopants \cite{Gotz1996,Zolper1996}. %
The measured values of $BE_{\text{Ga }2p3/2}$ for GaN$(1\bar{1}00)$ and GaN$(0001)$ after each cleaning step are shown in Table\,\ref{tab:sumup} together with the deduced surface BB. The full band structure is depicted in Fig.\,\ref{fig:Diagram} for the GaN$(1\bar{1}00)$ case. Remarkably, the cleaning steps drastically affect the surface BB and similar trends are identified for the two surface orientations. Specifically, the \textit{HCl etching} results in a flat band condition for GaN$(0001)$ and in a downward BB for GaN$(1\bar{1}00)$, the \textit{annealing} induces a moderate $0.1$--$0.2$\,eV upward BB, which is further increased to $0.5$--$0.6$\,eV after the \textit{O$_2$ plasma} for both orientations. Similar BB trends have been reported for HCl and UV/O$_3$ exposed GaN$(0001)$, although absolute BB values were not provided \cite{King1998,Tracy2003,Shalish2001,Rickert2002,Uhlrich2008}. Note that these surface electronic properties, hence, differ from those of an ideal atomically clean GaN surface \cite{Kocan2002,Rosa2006,Himmerlich2014}. The formation of the upward BB corresponds to the ionization of the donors in the bulk of the n-type GaN layer due to the localization of their electrons at the surface. It can be driven by either the presence of empty states in the GaO$_x$ layer at the surface, which are located below the GaN CBM \cite{Miao2010}, or a charge transfer process toward adsorbates induced by surface polarization (\eg{}, polarized Ga-OH and Ga-O-Ga bonds) or redox reactions \cite{Wang2017}. The observed $0.2$\,eV downward BB after \textit{HCl etching} for the GaN$(1\bar{1}00)$ surface essentially means flat band conditions in the n-type GaN layer with the addition of a thin electron accumulation layer within the CB at the surface, similarly to the InN$(0001)$ case \cite{Mahboob2004}.
The usual statement that HCl removes surface states within the GaN band gap \cite{Martinez2000,Shalish2001,Lee2006,Chen2007,Uhlrich2008,Gurwitz2011} cannot solely account for the formation of such an accumulation layer. The formation of intrinsic surface donors located above the GaN CBM or a charge transfer from adsorbates toward the GaN layer \cite{Schultz2016} is required. The formation of N vacancies acting as a donor \cite{Neugebauer1994} has been suggested in Refs.\,\cite{Sun2000,Rickert2002} but recent calculations are showing that such a donor state is located well below the GaN CBM when properly scaling the GaN band gap \cite{Miceli2015}. Adsorbates such as $\cdot$ClO$_n$H$_m$ can be considered since their removal by the \textit{N$_2$ annealing} causes the disappearance of the electron accumulation layer. H adsorbates may also have a dominant contribution. Indeed, a decrease of the GaN$(1\bar{1}00)$ and $(0001)$ surface potentials has been reported when decreasing the pH (\ie{} increasing the H$^+$ concentration) of a solution in contact with these surfaces\,\cite{Steinhoff2003,Wallys2012,Anvari2018}. Along the same line, H atoms have been reported to passivate the surface states of an atomically clean GaN$(1\bar{1}00)$ surface and to desorb at high temperature \cite{Lymperakis2017}. Therefore, we propose that $\cdot$ClO$_n$H$_m$ and/or $\cdot$H terminations at the GaN$(1\bar{1}00)$ and $(0001)$ surfaces act as surface donors. Overall, as shown in Table\,\ref{tab:sumup}, it is clear that the three cleaning steps investigated here can be used to tailor the GaN surface BB over several hundreds of meV, as qualitatively intended in other reports \cite{Gurwitz2011,Laehnemann2016a}.

\subsection{Work function}
The impact of the cleaning steps on the GaN WF is further investigated in vacuum by XPS and in N$_2$ atmosphere by KPFM. All results are summarized in Table\,\ref{tab:sumup}. For the two GaN orientations, the applied surface treatments modulate the GaN WF in a range larger than $1.4$\,eV in vacuum and $0.6$\,eV in N$_2$ atmosphere. The WF span surpasses the one of the BB, hence evidencing the existence of a surface dipole whose amplitude is modulated by the various adsorbates brought and removed by the cleaning steps. The estimation of the absolute surface dipole amplitude requires the knowledge of the electron affinity measured at a pristine GaN surface free of any surface dipole and surface reconstruction, which is in practice difficult to reach\,\cite{Kahn2016}. A value of $4.06$\,eV is proposed in Ref.\,\cite{Portz2018} based on scanning tunneling spectroscopy performed in high vacuum on a atomically cleaned GaN$(1\bar{1}00)$ surface. This value still stands as an approximation since the GaN$(1\bar{1}00)$ features a moderate surface reconstruction\,\cite{Lymperakis2013}. Nevertheless, we further use it here to estimate the dipole amplitude $\delta$ of the cleaned GaN$(1\bar{1}00)$ and $(0001)$ surfaces. The obtained values are reported in Table\,\ref{tab:sumup}. Through the consecutive cleaning steps carried on the GaN$(1\bar{1}00)$ surface, only the \textit{O$_2$ plasma} builds up a large surface dipole ($0.7$\,eV), pointing inward the GaN. In the GaN$(0001)$ case, an $0.8$\,eV dipole is observed after \textit{HCl etching}. This dipole is removed by \textit{N$_2$ annealing} and recovered after \textit{O$_2$ plasma} exposure. The different dipoles likely include electrostatic and chemical contributions. The first one occurs as a result of charge transfer between the adsorbate layer and the GaN surface. The second contribution relates to the covalent bonding of adsorbates to the GaN, which deeply modifies the chemical nature of the nitride surface. For a detailed evaluation of the microscopic origin of each of these dipoles, information on the surface bond configuration would thus be required, which lies out of the scope of the present study. 

We note that, for a given surface and cleaning step, different WFs are measured in vacuum by XPS and in N$_2$ atmosphere by KPFM. This discrepancy is attributed to the presence of a different amount of physisorbed species (\eg{}, water) on the surface depending on the measurement environment.

\subsection{UV light illumination}
Since UV light exposure has been largely reported to affect the photoluminescence \cite{Behm2000,Foussekis2009,Pfuller2010,Hetzl2018,Rousseau2018} and transport \cite{Gurwitz2011,Sanford2013,Posada2013} properties of GaN, we further investigate the change in the electronic properties of the cleaned GaN$(1\bar{1}00)$ and $(0001)$ surfaces when exposed to the light of a LED emitting at a wavelength of $365$\,nm. The measured core levels, VBM and SEC under illumination are shown in Fig.\,\ref{fig:ligthdependence} for the cleaned GaN$(1\bar{1}00)$ surface. Similar trends are observed for the cleaned GaN$(0001)$ surface. Especially, no shift of the core levels with respect to the dark state is observed, even for surfaces featuring a large upward BB such as after O$_2$ plasma exposure. It evidences the absence of a photo-induced screening of the BB, which differs from previous reports of Sezen \etal{Sezen2011,Sezen2014}. These authors have observed a Ga\,$2p_{3/2}$ core level shift of up to $0.15$\,eV when illuminating with a $405$\,nm laser an \insitu{} sputter-cleaned GaN$(0001)$ surface. The difference is attributed to the lower excitation density provided in our case by the UV LED and to a lower density of surface states, since both can affect the amplitude of the surface photovoltage\,\cite{Aphek1998}.

The dependence of the WF on the illumination is further scrutinized. In the absence of band bending screening, the photo-induced modification of the WF corresponds to a change in the surface dipole labeled $\Delta \delta_\text{LED}$ and defined as $WF_\text{LED} - WF_\text{dark}$. The obtained values of $\Delta \delta_\text{LED}$ measured in vacuum by XPS and in N$_2$ atmosphere by KPFM are reported in Table\,\ref{tab:sumup} for the two GaN surfaces after the consecutive application of the different cleaning steps. Remarkably, $\Delta \delta_\text{LED}$ in vacuum reaches $-0.2$\,eV after \textit{HCl etching} for both GaN orientations and is negligible in the other cases. Analogously, in N$_2$ atmosphere, $\Delta \delta_{LED}$ has a larger amplitude after \textit{HCl etching} compared to after the other surface treatments. It evidences a photo-dependence of the surface dipole built up by the $\cdot$ClO$_n$H$_m$ termination.
Zhuang \etal{Zhuang2017} have shown that upon UV light exposure, Cl adatoms on graphene can be excited, resulting in various chemical reactions. However, such process occurs only for deep UV light ($\lambda < 344$\,nm), which does not fit our case. Further understanding of the observed photo-dependence would thus require a further microscopic description of the involved surface dipole.

In conclusion, UV light can modify the GaN WF (i) by screening the band bending, and (ii) by perturbing the surface dipole. A priori, these two contributions cannot be disentangled when performing only surface photovoltage measurements by Kelvin probe. On this basis, one should revise the conclusions drawn in Refs\,\cite{Shalish2001,Gurwitz2011,Foussekis2012,McNamara2013,Winnerl2017} where surface photovoltage measurements carried on HCl treated GaN are discussed only in terms of BB screening.


\section{Summary and Conclusions}
We have investigated the electronic properties of GaN$(1\bar{1}00)$ and $(0001)$ surfaces after three common cleaning steps. Namely, HCl etching, $400$\degC{} annealing in N$_2$ atmosphere, and O$_2$ plasma exposure. All these methods are suitable to prepare two-dimensional and nanostructured films and devices. The nature of the adsorbates remaining on the GaN surfaces after the various cleaning steps is scrutinized and found to be similar for the $(1\bar{1}00)$ and $(0001)$ orientations. The GaN BB and WF are further quantified. They exhibit a large dependence on the cleaning steps as a result of both the modification of surface states and the creation of distinct surface dipoles due to the presence of different adsorbates. Interestingly, a downward BB is obtained in GaN$(1\bar{1}00)$ after HCl etching, revealing a charge transfer process from surface donors toward the GaN layer.
For all investigated surfaces, UV-A illumination does not screen the surface BB. However, the same illumination largely modifies the amplitude of the surface dipole existing on GaN surfaces treated with HCl. This observation disqualifies the sole use of surface photovoltage measurements by Kelvin probe to characterize the GaN surface BB. 
In general, we recommend the use of the cleaning steps discussed here to predictively tune the electronic properties of air-exposed n-type GaN surfaces (surface BB and light-dependent surface dipole), which otherwise can drastically and erratically change due to the presence of uncontrolled surface contaminants. \\

\section{acknowledgments}
We thank Katrin Morgenroth, Carsten Stemmler and Hans-Peter Sch\"onherr for their dedicated maintenance of the molecular beam epitaxy system and Sander Rauwerdink for his support with the sample treatments. We are indebted to Wolfram Jaegermann for fruitful discussions and to Henning Riechert for his continuous encouragement and support. Furthermore, we thank Martin Heilmann for a critical reading of the manuscript. Funding from the Bundesministerium für Bildung und Forschung through project FKZ:13N13662, 13N13656, 13N13657 and, 13N13658 are gratefully acknowledged. Sergio Fernández-Garrido acknowledges the partial financial support received through the Spanish program Ramón y Cajal (co-financed by the European Social Fund) under grant RYC-2016-19509 from Ministerio de Ciencia, Innovación y Universidades.

\bibliography{BareGaNsurfaceproperties}

\end{document}